\documentstyle[preprint,aps]{revtex}
\preprint {IMSc-99/12/40}
\begin{document}
\tightenlines
\draft
\title{Edge magnetoplasmon excitations in a Quantum Dot in high magnetic
fields}
\author{Subhasis Sinha}

\address
{The Institute of Mathematical Sciences, Madras 600 113, India.}
\date{\today}
\maketitle

\begin{abstract}

We investigate the collective magnetoplasmon excitations in a quantum dot
containing finite number of electrons in the high magnetic field limit. We
consider the electrons in the lowest Landau level and neglect mixing
between the higher Landau levels. The dispersion relation of these edge
modes are estimated following the energy weighted sum-rule approach. In
this finite size system the edge magnetoplasmon modes have different
multipolarities (or angular momemtum $l$). Their dependence on the
magnetic field and on the system size is investigated. With increasing
magnetic field, energy of these collective modes decreases and in the bulk
limit they become gapless. We also consider the breathing mode of a dot
in the presence of a strong magnetic field, and the energy of this mode
approaches the cyclotron frequency $\hbar \omega_{c}$.  
\end{abstract}


\narrowtext
Recent developments in nanofabrication technology have made it possible
to manufacture electronic systems with reduced dimensionality. Quantum
dot is one such nanosystem, which is very interesting both
experimentally and theoretically. A quantum dot is a two-dimensional
electron system where the electrons are confined within a finite area by
applying a gate voltage\cite{chakraborty}.
It is an example of a finite size quantum system where the number of
electrons can be varied from a few electrons to a few thousand electrons.
This nanosystem
shows many interesting properties in the presence of a magnetic field.
One of the interesting properties in the presence of strong magnetic 
field is the chiral edge magnetoplasmon excitations. The edge modes in the
quantum Hall bar have been studied theoretically using one dimensional
field theory and numerical diagonalisation of the few particle system
\cite{wen,stone}.  
The edge modes of a two-dimensional electron system in the presence of
positive ion background have
been studied by several authors \cite{stringari,lipparini}. Giovanazzi et
al \cite{stringari} have calculated the dispersion relation of the edge
modes 
of the electron gas confined in a jellium disk in the long wave-length
limit.
Recently collective excitations in the finite size nanostructures like
quantum dot have become very interesting, because
many device parameters can be controlled by adjusting the gate voltages.
The number of electrons in the system can also be varied. In particular, 
the dipole
excitation in a quantum dot with parabolic confinement, and in the
presence of a strong magnetic field has been discussed in ref.
\cite{maksym}. Evidence has been found for a strong collective dipole mode
in the presence of a magnetic field in far infrared spectroscopy
experiment \cite{merket,demel}. Recently the collective charge and spin
excitation in a dot containing about 200 electrons has been observed
experimentally \cite{heitmann}. Non parabolic confinement has been
observed in this experimental device, but the exact nature of the
confinement is not well understood. The edge magneto-plasmon modes of a
quantum dot with
finite number of electrons have been studied by using
magneto-hydrodynamics
by several authors \cite{shikin,fetter,peteers,zeremba}. 

The collective excitations in the finite fermion systems like nuclei and
metal clusters have been extensively studied within random phase
approximation(RPA) and using sum-rule approach in the last years by
several
authors \cite{lipparini2,brack}. This method has also been
applied to analyze the multipole excitations in the two dimensional
quantum dot and antidot systems \cite{lipparini,sinha,serra}.
Most of the theoretical works have been done numerically or by using
classical magneto-hydrodynamics. The main aim of this work is to estimate
the collective modes of a dot in the presence of a strong magnetic field 
microscopically, and also to obtain analytical results for the
dispersion relations of these modes.

In this paper we study the low-lying multipole excitations and the
breathing
mode of a quantum dot
in the presence of a strong magnetic field by using RPA sum-rule
approach.
We consider the electrons in the lowest Landau level, and derive a
simple semiclassical energy functional for the electrons. By minimising
the energy functional we obtain the ground state density profile of the
electrons. Then we generalise the sum-rules in the presence of a magnetic
field. Different moments of the RPA strength distribution function are
calculated by using the ground state density profile. Finally we obtain
the analytic expressions for the dispersion relation of the low-lying
magnetoplasmon modes with different multi-polarities in a quantum dot.
With increasing magnetic field, the energy of these modes decreases and
in the bulk limit these modes become gap-less. Neglecting the $1/N$
corrections, the dispersion relation takes a very general form, where only
one parameter contains the information about the two-body interaction and
the shape of the density profile. Also, the $1/N$ corrections
of these modes are calculated. In the strong magnetic field limit the
dispersion relation of these modes agrees with the dispersion relation
obtained from hydrodynamics\cite{shikin}. We also calculate the energy of
breathing mode excitation, which is much larger than the energy of
low-lying edge modes. We obtained the most general form of the dispersion
relation of the edge excitations in strong magnetic field limit. We
also calculated the finite size $N$ dependent corrections of the energy of
these modes. Analyzing the dispersion relation in a strong magnetic field,
we have shown the stability of the self-consistent semicircular density
profile, which shows that this semi circular density is a better ansatz
for ground state density than commonly used flat density profile.    

The hamiltonian of the two-dimensional electron system in the presence of
a magnetic field in the perpendicular direction is given by,
\begin{eqnarray}
H = & & \sum_{i}\frac{1}{2m^{*}}[(\hat{p}_{xi} + \frac{1}{2}m^{*}
\omega_{c}
y_{i})^{2}
+(\hat{p}_{yi} - \frac{1}{2}m^{*} \omega_{c} x_{i})^{2}] +
\sum_{i}\frac{1}{2}m^{*}\omega_{0}^{2} r_{i}^{2}\nonumber \\& & +
\sum_{i<j}
V(|\vec{r}_{i} -\vec{r}_{j}|), 
\end{eqnarray}
where $m^{*}$ is the effective mass, $\omega_{0}$ is the frequency of the
external parabolic confinement, and $\omega_{c}$ is the cyclotron
frequency $\frac{e B}{m^{*} c}$.  
Neglecting the two-body interaction, the single particle Hamiltonian can
be solved exactly. The single-particle energy levels are,
\begin{equation}
E_{n,m} = (2 n + |m| + 1)\Omega - \frac{m}{2}\omega_{c},
\end{equation}
where $\Omega = \sqrt{\omega_{0}^{2} + \frac{\omega_{c}^{2}}{4}}$, and
$m$ is the angular momentum quantum number, and $n$ denotes Landau level
index. In the presence of a strong magnetic field all electrons occupy
the lowest Landau level(LLL), energy levels become almost degenarate and
they form a band. In the lowest Landau level the single particle wave
functions are,
\begin{equation}
\psi_{m}(\vec{r}) = \frac{1}{\sqrt{\pi l_{0}^{2}}}
\frac{1}{\sqrt{m!}}(\frac{z}{l_{0}})^{m} e^{-r^{2} / 2 l_{0}^{2}},
\end{equation}
where, $l_{0} = \sqrt{\frac{\hbar}{m^{*} \Omega}}$, and $z = x+iy$ is the
complex coordinate.
In the high magnetic field limit the effect of higher Landau levels can be
neglected and the many body wave function can be constructed out of these
lowest Landau level basis. For electrons at the
filling factor $\nu = 1$ the Laughlin wave function can be obtained by
constructing the Van-der Monde determinant of the above single particle
states,
\begin{equation}
\psi(\vec{r}_{1},\vec{r}_{2},....\vec{r}_{n}) = \Pi_{i< j} (z_{i} - z_{j})
e^{-\sum_{i} z_{i}\bar{z}_{i} / 2 l_{0}^{2}}.
\end{equation}
In general we can write the ground state many body wave function
constructed out of the LLL as,
\begin{equation}
\psi(\vec{r}_{1},\vec{r}_{2},..\vec{r}_{i}..) = f(z_{1},
z_{2},..z_{i},..) e^{-\sum_{i} z_{i}\bar{z}_{i} / 2 l_{0}^{2}}.
\end{equation}
We can calculate the energy functional in the lowest Landau level, by
calculating the expectation value of the Hamiltonian with respect to the
ground state wave function given above. After doing some algebra and using
some properties of the Bargman space \cite{girvin}, we arrive at the
following energy functional,
\begin{eqnarray}
E[\psi]  & = & \frac{1}{2}\hbar\omega_{c} \int |\psi|^{2} d^{2} r + m^{*}
\Omega (\Omega - \omega_{c} / 2) \int r^{2} |\psi|^{2} d^{2} r \nonumber\\ 
& & +
\frac{1}{2}\int d^{2} r_{1} \int d^{2}r_{2} V(|\vec{r}_{1} - \vec{r}_{2}|)
|\rho(r_{1}, r_{2})|^{2} d^{2} r_{2},
\end{eqnarray}
where,
\begin{equation}
|\rho(R_{1}, R_{2})|^{2} = <\vec{r}_{1},
\vec{r}_{2},...\vec{r}_{i},..|\sum_{ij} \delta(\vec{R}_{1} - \vec{r}_{i})
\delta(\vec{R}_{2} - \vec{r}_{j}) |\vec{r}_{1},
\vec{r}_{2},...\vec{r}_{i}...> . 
\end{equation}
Within Hartree approximation we can write down a simple 
local density functional for lowest Landau level,
\begin{eqnarray}
E[\rho] & = & \int d^{2} r [\frac{1}{2}\hbar \omega_{c} + m^{*} \Omega
(\Omega - \frac{\omega_{c}}{2}) r^{2} ] \rho(r)\nonumber \\ & & +
\frac{1}{2}
\int d^{2}r_{1} \int
d^{2} r_{2} \rho(r_{1}) V(|r_{1} - r_{2}|) \rho(r_{2}).
\end{eqnarray}
In the case of a flat droplet, the Hartree energy typically goes as $\sim
N^{3/2}$, and the exchange correlation energy goes as $\sim N$, where $N$
is the number of particles in the droplet\cite{mcdonald}. Hence in the
large $N$ limit, the
exchange term may be neglected and we obtain the above simple looking
energy
functional.
We minimize the free energy $E - \mu N$ with respect to $\rho$,
keeping the total number of particles $N $ fixed,
and obtain the following integral equation for the density,
\begin{equation}
m^{*} \Omega (\Omega - \omega_{c}/2) r^{2}  + \int d^{2} r' V(|\vec{r} -
\vec{r'}|) \rho(r') = \mu - \frac{1}{2} \hbar \omega_{c}.
\end{equation}
For coulomb interaction, $V(r) = e^2 /r$ the solution of the above
integral
equation is \cite{shikin,lieb},
\begin{equation}
\rho(r) = n_{0} \sqrt{1 - (\frac{r}{R})^{2}},
\end{equation}
where, $n_{0} = \frac{3N}{2 \pi R^{2}}$, and $R^{3} = \frac{3 e^{2} \pi
N}{8 m^{*} \Omega (\Omega - \frac{\omega_{c}}{2})}$. From this simple
energy functional we obtain the nonperturbative density of the electrons.

We now consider the edge magnetoplasmon excitations in the dot.
Magnetoplasmon excitation are the multipole excitations of a quantum dot
in the presence of a  magnetic field. We denote the multipolarity of the
operator by the
number $k$. The excitation with multipolarity
$k$ is
generated by the excitation operator $F = z^{k}$, where $z$ is the complex 
coordinate $x + i y$. Excitation energies of the above modes can be
estimated by sum-rule method. Many useful quantities may be calculated
from the strength function, that is defined as
\begin{equation}
S_{\pm}(E) = \sum_{n} |<n| F_{\pm} |0>|^{2} \delta(E - E_{n}),
\end{equation}
where, $E_{n}$, and $|n>$ are the excitation energy and excited state
respectively, and $F_{+} = F$, and $F_{-} = F^{\dagger}$. Various
energy-weighted sum
rules can be derived through the moments of the strength function, and are
written below,
\begin{eqnarray}
m_{k}^{\pm} & = & \frac{1}{2}\int E^{k} (S_{+}(E) \pm S_{-}(E)) dE
\nonumber\\
& = & \frac{1}{2}(<0|F^{\dagger}(\hat{H} - E_{0})^{k}F|0> \pm <0|F
(\hat{H} - E_{0})^{k} F^{\dagger}|0>).
\end{eqnarray}
Some useful moments can be written in terms of the commutators of the
excitation operator $F$ with the many body Hamiltonian $H$, and they are
given below,
\begin{eqnarray}
m_{0}^{-} & = & \frac{1}{2}<0|[F^{\dagger}, F]|0>,\\
m_{1}^{+} & = & \frac{1}{2} <0|[F^{\dagger}, [H, F]]|0>,\\ 
m_{2}^{-} & = & \frac{1}{2}<0|[J^{\dagger}, J]|0>\\
m_{3}^{+} & = & \frac{1}{2}<0|[J^{\dagger}, [H, J]]|0>,
\end{eqnarray}
where $J = [H, F]$. In the presence of magnetic field $+k$ and $-k$
collective modes split, and the corresponding strength distributions are
sharply peaked at these collective frequencies. Near the collective
excitation energy, we can approximate the strength distribution by delta
function, $S_{\pm} = \sigma_{\pm} \delta(E - E_{c\mp})$. Since for
the multipole modes $m_{0}^{-} = 0$, we obtain $\sigma_{+} = \sigma_{-}$.
Within this approximation the low lying multipole modes can be estimated
by using the above moments of the strength distribution function and can
be written as,
\begin{equation}
E_{c \mp} = \sqrt{\frac{m_{3}^{+}}{m_{1}^{+}} - \frac{3}{4}
(\frac{m_{2}^{-}}{m_{1}^{+}})^{2}} \pm \frac{m_{2}^{-}}{2},
\end{equation}
where, $\pm$ sign correspond to $k$ and $-k$ multipolarities.
Similar expression for the excitation frequencies of the edge multipole
modes can be derived by using the variational principle.  
Given the $N$ electron ground state $|0>$, it is possible to find the
collective excitation energy and the collective state $|c>$, if one able
to find an operator $O^{\dagger}$, which satisfies the following equation
of motion,
\begin{equation}
[\hat{H}, O^{\dagger}] = \hbar \omega_{coll} O^{\dagger}.
\end{equation}
The state $O^{\dagger}|0>$ has the excitation energy $\hbar
\omega_{coll}$.
This excitation energy is obtained from,
\begin{equation}
\hbar \omega_{coll} = \frac{<0|[O, [\hat{H}, O^{\dagger}]]|0>}{<0|[O,
O^{\dagger}]|0>}.
\end{equation}
For $k = 1$ the dipole excitation operator is,
\begin{equation}
O^{\dagger} = \frac{1}{2} \sum_{i} (z_{i} - \frac{i}{m^{*} \Omega}
\hat{p}_{i+}),
\end{equation}
where $p_{+} = p_{x} + i p_{y}$. Similarly, we may take the
variational
ansatz for the higher multipole excitations in the following form,
\begin{equation}
O^{\dagger} = F + a J~~~;~~~J = [H, F],
\end{equation}
where $a$ is the variational parameter, $F = \sum_{i} z_{i}^{k}$, and $J =
(\frac{-i \hbar k}{m^{*}})
\sum_{i} z_{i}^{(k - 1)} \hat{D}_{+}$,
where, $\hat{D}_{+} = \hat{p}_{+} + \frac{i m^{*} \omega_{c}}{2} z$.
The collective excitation energy in terms of the the energy weighted sum
rules is,
\begin{equation}
\hbar \omega_{coll} = \frac{m_{1}^{+} + 2 a m_{2}^{-} + a^{2} m_{3}^{+}}{2
a m_{1}^{+} + a^{2} m_{2}^{-}}.
\end{equation}
After minimising the above expression with respect to the variational
parameter $a$, the following expression for the collective excitation
frequency is obtained,
\begin{equation}
\hbar \omega_{coll} = \sqrt{m_{3}^{+} - \frac{3}{4}
(\frac{m_{2}^{-}}{m_{1}^{+}})^{2}} + \frac{m_{2}^{-}}{2}.
\end{equation}
This agrees with the previous expression in eqn.(17), obtained by the
approximate form the strength distribution.

Now we explicitly evaluate the the sum-rules by calculating the
commutators. After doing some algebra the sum-rules can be written in the
following form. The first energy weighted moment $m^{+}_{1}$ is given by,
\begin{equation}
m_{1}^{+}(k)  =   \frac{\hbar^{2} k^{2}}{m^{*}} < r^{2(k - 1)}>, 
\end{equation}
where $ <...>$ denotes the average weighted with the ground state density.
The contribution for the third moment $m^{+}_{3}$ coming from the kinetic
energy of the hamiltonian $m^{+}_{3}(T)$ can be written as,
\begin{eqnarray}
m_{3}^{+}(T) & = & (\frac{\hbar^{2} k^{2}}{m^{*3}}) [ \hbar^{2} k (k - 1)
\int d^{2} r r^{2(k - 2)} |\nabla \psi|^{2} \nonumber\\ & & + 2 \hbar^{2}
(k- 1) (k - 2)
< r^{2(k - 3)} \hat{l}_{z}^{2} > - 3 \hbar^{2} k (k - 1) m^{*} \omega_{c}
< r^{2(k - 2)} \hat{l}_{z} >\nonumber\\ & & +  \hbar^{2} m^{*2}
\omega_{c}^{2}(\frac{3}{4}k^{2} + \frac{1}{4} k) < r^{2 (k - 1)} >.
\end{eqnarray}
Similarly the contribution of the external potential term in $m^{+}_{3}$
is,
\begin{equation}
m_{3}^{+}(V)  =  (\frac{\hbar^{2} k^{2}}{m^{*3}}) k m^{*2} \omega_{0}^{2}
< r^{2 (k - 1)} >.
\end{equation}
The most important contribution comes from the electron-electron
interaction term, and is given by,
\begin{eqnarray}
m_{3}^{+}(ee) & = & \frac{\hbar^{4} k^{2}}{2 m^{*2}}[ \int d^{2}r
\nabla^{2} V_{H}(r) r^{2 (k - 1)} \rho(r) + 2 (k - 1) \int d^{2} r
\frac{\partial V_{H}}{\partial r} r^{2 (k - 3)} \rho(r) \nonumber \\
& & + \int d^{2}r \int d^{2}r' \rho'(r) e^{-i l \theta} \frac{1}{|\vec{r}
- \vec{r'}|} e^{i l \theta^{'}} r'^{(k - 1)} \rho'(r') ],
\end{eqnarray}
where $V_{H}(r)$ is the Hartree potential.
In the presence of a magnetic field there is nonvanishing second moment
$m^{-}_{2}$,
\begin{equation}
m_{2}^{-}  =  \frac{\hbar^{2} k^{2}}{m^{*2}} [ - \hbar m^{*} \omega_{c}
k < r^{2 (k - 1)} > + 2 \hbar (k - 1) < r^{2 (k - 2)} \hat{l}_{z} >],
\end{equation}
where $\hat{l}_{z}$ is the angular momentum operator with eigenvalue $m$.
We now consider the limit in which the magnetic field is large, so that
all electrons are in the lowest landau level. Neglecting the mixing of
higher
Landau levels, if we consider the wave function in the following form,
\begin{equation}
\psi({z_{i}}) = f(z_{1}, z_{2},....z_{i},...) e^{- \sum z_{i} \bar{z}_{i}
/2
l_{0}^{2}},
\end{equation}
then the above sum-rules can be simplified and written in terms of moments
of radius $r$.
In the lowest Landau level the sum-rules $m_{3}^{+}$ and $m_{2}^{-}$ can
be written as,
\begin{eqnarray}
& & m_{3}^{+}(T) + m_{3}^{+}(V) = \frac{\hbar^{2} k^{2}}{m^{*3}} [
\hbar^{2} m^{*2} \Omega^{2} ({3 ( (k - 1) -
\frac{k}{2}(\frac{\omega_{c}}{\Omega}))^{2}} + 1) < r^{2 (k - 1)} >
\nonumber \\
& & - 3 \hbar^{3} m^{*} \Omega (k - 1)^{2} (2 (k - 2) -
k(\frac{\omega_{c}}{\Omega})) < r^{2 (k - 2)} > \nonumber \\ & & + 4
\hbar^{2} (k - 1)^{2}
(k - 2)^{2} < r^{2 (k - 3)} >], \\
& & m_{2}^{-} = \frac{\hbar^{2} k^{2}}{m^{*2}}[-\hbar m^{*} \omega_{c} k <
r^{2 (k - 1)} > \nonumber \\
& & + 2 \hbar (k - 1) (m^{*} \Omega < r^{2 (k - 1)} > - \hbar (k - 1) 
< r^{2 (k - 2)} >)].
\end{eqnarray}
We can write down the expressions for the ratios of moments in the
following form,
\begin{eqnarray}
\frac{m_{2}^{-}}{m_{1}^{+}} & = & 2 (k - 1) (\Omega -
\frac{\omega_{c}}{2}) - \omega_{c} + \frac{\alpha_{1}}{R^{2}}, \\
\frac{m_{3}^{+}(T) + m_{3}^{+}(V)}{m_{1}^{+}} & = & \hbar^{2} \Omega^{2}
[1 + 3 (k - 1 - \frac{k}{2}(\frac{\omega_{c}}{\Omega}))^{2}] +
\frac{\alpha_{2}}{R^{2}} + \frac{\beta_{2}}{R^{4}}.
\end{eqnarray}
For large number of electrons, if we neglect the terms $O(1/R^{2})$ and
the
higher order terms, then the excitation energies can be written in the
most general form, given below,
\begin{equation}
E_{coll}(k) = (k - 1) \hbar (\Omega - \frac{\omega_{c}}{2}) +
\sqrt{\hbar^{2}\Omega^{2}
+ \Delta(e)} - \frac{\omega_{c}}{2}.
\end{equation}
The coefficients $\alpha_{1}$, $\alpha_{2}$, $\beta_{2}$ depend on 
the density of the electrons, and the function $\Delta(e)$ depends on the
nature of the two body interaction
and also on the shape of the density profile. 
Now we can estimate the low-lying multipole excitations by using the
semi-circular
density profile. 
The parameters in the above form of the collective frequency can be
evaluated by using the semicircular density profile $ \rho = n_{0}
\sqrt{1 - (r/R)^{2}}$, 
\begin{eqnarray}
\Delta(e) & = & 2 \hbar^{2} \Omega (\Omega - \frac{\omega_{c}}{2})
[\frac{\Gamma(k + \frac{1}{2})}{\Gamma(k) \Gamma(3/2)} - k], \\
\alpha_{1} & = & \hbar \Omega (k - 1) (2 k + 1), \\
\alpha_{2} & = & - \frac{3}{2} \hbar^{2} \Omega^{2} (k - 1) ( 2 k + 1) (
2 (k - 2) - k (\frac{\omega_{c}}{\Omega})),\\
\beta_{2} & = & \hbar^{2} \Omega^{2} ( k - 1)( k - 2) ( 2k + 1) ( 2k - 1).
\end{eqnarray}
Since the semi-circular density profile is the exact solution of the
Hartree
equation, the dispersion relation of the edge modes obtained by using
this density
is a non-perturbative result. Further the asymptotic results do not
contain
the coupling constant $e^{2}/l_{0}$, only the finite size corrections
depend on the coupling constant. This density profile goes to zero at
the turning point $R$ and the difussive tail is absent in this case. But
the
difussive tail in the exact density distribution can only contribute $1/N$
corrections to the spectrum of low-lying excitations \cite{sinha}. From
the energy levels of the non-interacting electrons, we can estimate the
magnetic field above which all $N$ particles go into the lowest Landau
level. This gives the condition $2 \hbar \Omega > (N - 1) (\Omega -
\omega_{c}/2)$(or $\omega_{0}/\omega_{c} < 1/\sqrt{N}$). If $k<< N$,
which is true for the low-lying modes, then in the strong field limit we
can expand the expression for collective modes in terms of
$\Delta(e)/\hbar^{2} \Omega^{2}$, and obtain the following result in
strong magnetic field limit 
\begin{equation}
E_{c}(k) = \omega_{0} \frac{\omega_{0}}{\omega_{c}} \frac{\Gamma(k +
1/2)}{\Gamma(k) \Gamma(3/2)} + O((\frac{\omega_{0}}{\omega_{c}})^4).
\end{equation}
This result now be compared with the dispersion relation obtained by,
Shikin et al \cite{shikin} using classical hydrodynamics,
\begin{eqnarray}
\omega_{k}^{-} & = & \sqrt{k \Omega_{kk}^{2} + \omega_{c}^{2}/4}
-\omega_{c}/2, \\
k \Omega_{kk}^{2} & = & \omega_{0}^{2} \frac{\Gamma(k + 1/2)}{\Gamma(k)
\Gamma(3/2)},
\end{eqnarray}
where $\hbar \omega^{-}_{k}$ is the same as $E_{c}(k)$ in our notation.
If we expand this expression in terms of $\frac{\omega_{0}}{\omega_{c}}$,
the
leading term agrees with the dispersion relation obtained from sum-rule
approach. 

For comparison, we may also estimate the edge excitations in a quantum dot
by using the
density profile of the non-interacting electrons at the filling factor 
$\nu = 1$. In the large $ N $ limit, the density profile of the system can
be approximated by, $ \rho_{0} \theta( R - r)$, where, $\rho_{0} =
\frac{1}{\pi l_{0}^{2}}$. This density profile is very sharp near the
edge, and therefore singularities arise in evaluating $\Delta$, but after
doin the integrals and then taking the limit($r \rightarrow R$) at the
edge, the singularities cancel out and we obtain finite value of the
parameter $\Delta$. The parameter $\Delta$ in this case can be derived as,
\begin{equation}
\Delta(e) = 2 \hbar \Omega \frac{e^{2}}{l_{0}} \frac{k}{\sqrt{N}} [ 1 -
\sum_{m=1}^{k} \frac{1}{2 m - 1} ].
\end{equation}
This result agrees with the
value obtained by Giovanazzi et al \cite{stringari}.
In this case the excitation energies of various multipole modes vanishes
at different values of magnetic field. The approximate value of the
magnetic field,
where excitation energy of the $k$ th mode becomes negative is,
\begin{equation}   
(\frac{\omega_{0}}{\omega_{c}})^{3/2} < \frac{e^{2}}{\sqrt{2
N}\tilde{l}_{0} \hbar \omega_{0}}[\sum_{m = 1}^{k} \frac{1}{2m - 1} - 1],
\end{equation}
where $\tilde{l}_{0} = \sqrt{\frac{\hbar}{m^{*} \omega_{0}}}$. Energy of
the higher multipole modes vanish at lower magnetic field. For example 
the mode with multipolarity $k = 10$ becomes gapless at a magnetic
field $\sim 9.65T$, in a dot with $N = 40$ and $\hbar \omega_{0} = 5.4 
meV$. Softening of these edge modes indicate the edge reconstruction
of the dot and the formation of new ground state. This phenomena
indicates the instability of the flat density. The critical magnetic field
where the instability sets in is obtained from\cite{mcdonald,vignale},
\begin{equation}
(\frac{\omega_{0}}{\omega_{c}})^{3/2} \approx [ \frac{.5139
e^{2}}{\tilde{l}_{0} \hbar \omega_{0} \sqrt{N}}].
\end{equation}
Although $\Delta$ is negative for both the cases, the excitation energy
for low-lying modes
for the semi-circular density profile are positive and asymptotically go
to zero, which indicates the stability of the ground state. For comparisn,
we have shown in Fig.1 $\Delta$ for different multipolarities,
evaluated using two
different ground state densities. In Fig.2, the variation of few
low-lying modes with magnetic field is shown, using the dispersion law
given in eqn.(34) and eqn.(35). These modes vanish as $\sim 1/B$ with 
the increasing magnetic field. These low-lying modes are
important for the thermodynamics of the system at very low temperatures. 

Finally we consider the breathing mode of the dot within the same
formalism. The breathing mode is excitated by the excitation operater,
$ F = r^2$. The operator $J$ is,
\begin{equation}
J = [H , F] = (\frac{- i \hbar}{m^{*}})[\hat{p}_{x} x + x \hat{p}_{x} +
\hat{p}_{y} y + y \hat{p}_{y}].
\end{equation}
The first moment $m^{+}_{1}$ is given by,
\begin{equation}
m^{+}_{1} = \frac{2 \hbar^{2}}{m^{*}} < r^{2} >.
\end{equation}
The third moment can be written in the following way,
\begin{equation}
m^{+}_{3} = \frac{1}{2} \frac{\partial^{2}}{\partial \eta^{2}} < \eta |
\hat{H} | \eta > |_{\eta = 0},
\end{equation}
where $ | \eta > = e^{\eta J} | 0 >$. Now the moment $ m^{+}_{3}$ can be
evaluated by using the scaling properties of the wave function
\cite{bohigas},
\begin{equation}
e^{\eta J} | \psi > = e^{-2 \eta \hbar^{2} /m^{*}} | \psi(x e^{-2 \eta
\hbar^{2}/m^{*}} , y e^{-2 \eta \hbar^{2} / m^{*}}) >.
\end{equation}
From the above scaling property of the wave function, we obtain,
\begin{eqnarray}
m^{+}_{3} & = & 8 (\frac{\hbar^{2}}{m^{*}})^{2} [ < \hat{T} > +
\frac{1}{2} m^{*} \Omega^{2} < r^{2} > ] + 2
(\frac{\hbar^{2}}{m^{*}})^{2} E_{int} , \\
& = & 8 (\frac{\hbar^{2}}{m^{*}})^{2} [ m^{*} \Omega^{2} < r^{2} >  ] + 2
(\frac{\hbar^{2}}{m^{*}})^{2} E_{int},
\end{eqnarray}
where $ \hat{T} $ is the kinetic energy operator and $E_{int}$ is the
interaction energy of the system. In deriving the second step, we used the
properties of the lowest-Landau level wave functions. The general
expression of the excitation energy of the breathing mode is derived as,
\begin{equation}
E_{b} = \sqrt{ 4 \hbar^{2} \Omega^{2} + \frac{\hbar^{2}}{m^{*}}
\frac{E_{int}}{ < r^{2} >}}.
\end{equation}
Using the semi-circular density for the ground state of the electrons, we
obtain the 
non-perturbative result for the breathing mode,
\begin{equation}
E_{b} = 2 \hbar \Omega \sqrt{1 + \frac{1}{4}(1 - \frac{\omega_{c}}{2
\Omega})}.
\end{equation}
In the strong field limit this mode approaches the bulk collective mode
$\hbar \omega_{c}$.

To summerise, we have considered the quantum dot in the lowest Landau
level.
We have derived a simple local density functional for the electrons
in the LLL within the Hartree approximation. For the two-body coulomb
interaction                                   
a semicircular density profile of the electrons has been derived. We have
estimated the low-lyimg edge multipole modes within the sum-rule
approach.
Since the semi-circular density is exact within Hartree approximation,
we obtain the non-perturbative results for the dispersion realtions of the
collective modes by using this density profile.
The energy of these collective modes decreases with increasing magnetic
field. In the strong field
limit the most general expression for the dispersion relation of the edge 
modes has been derived and is given by, $ E(k) = \sqrt{\Omega^{2} +
\Delta(e)}
- \frac{\omega_{c}}{2} + (k - 1) \omega_{0}\frac{\omega_{0}}{\omega_{c}}$.
The parameter $\Delta(e)$ depends on the exact nature of two-body
interaction and on the number of electrons in the system. We evaluate the
parameter $\Delta(e)$ by using exact density, as well as by using the
density profile of the non interacting electrons at filling factor $\nu =
1$. In both cases $\Delta(e)$ is negative. In the case of flat density the
energy of the low-lying edge modes become zero at some magnetic field and
the softening of the edge modes indicate the instability of the ground
state. But for the semi-circular ground state density the low-lying modes
are positive and asymptotically go to zero, which shows the stability of
the self-consistent density profile. We have also
calculated
the breathing mode of a quantum dot, and in the strong magnetic field the
energy of the breathing mode approaches the cyclotron frequency
$\omega_{c}$. The main results of this paper are, the derivation of
the most general
dispersion relation of the edge modes in a quantum dot in a strong
magnetic field. The $1/N$ corrections of the
energy of these modes are also obtained. Stability of the self consistent
density profile has been shown by analysing the dispersion relation of the
low-lying edge modes. This result shows that the semicircular density is a
better ansatz for the ground state density than commonly used flat density
profile, in the strong magnetic field limit. These edge modes are
important for the edge excitation and edge reconstruction of quantum dot.
They also determine the low-temperature thermodynamic properties of the
quantum dot.

I would like to thank M. V. N Murthy for his helpful comments. I would
also like to thank Tapash Chakraborty and R. Shankar for critical reading
of the manuscript.

\newpage
\begin{figure}
\caption{The parameter $\Delta(e)$ in units of $\hbar^{2} \omega_{0}^{2}$,
for different multipolarities $k$, for a dot with $N= 50$,
$\frac{e^{2}}{\tilde{l}_{0} \hbar \omega_{0}} = 0.8$ and
$\omega_{c}/\omega_{0}
= 10$. The open triangles represent values calculated with
semi-circular density and the solid triangles denote the
same parameter for flat density profile.}

\caption
{The variation of four lowest multipole modes with magnetic field, in a
quantum dot with semi-circular density profile. Different values of $k$
denote different multipolarities.} 

\end{figure}

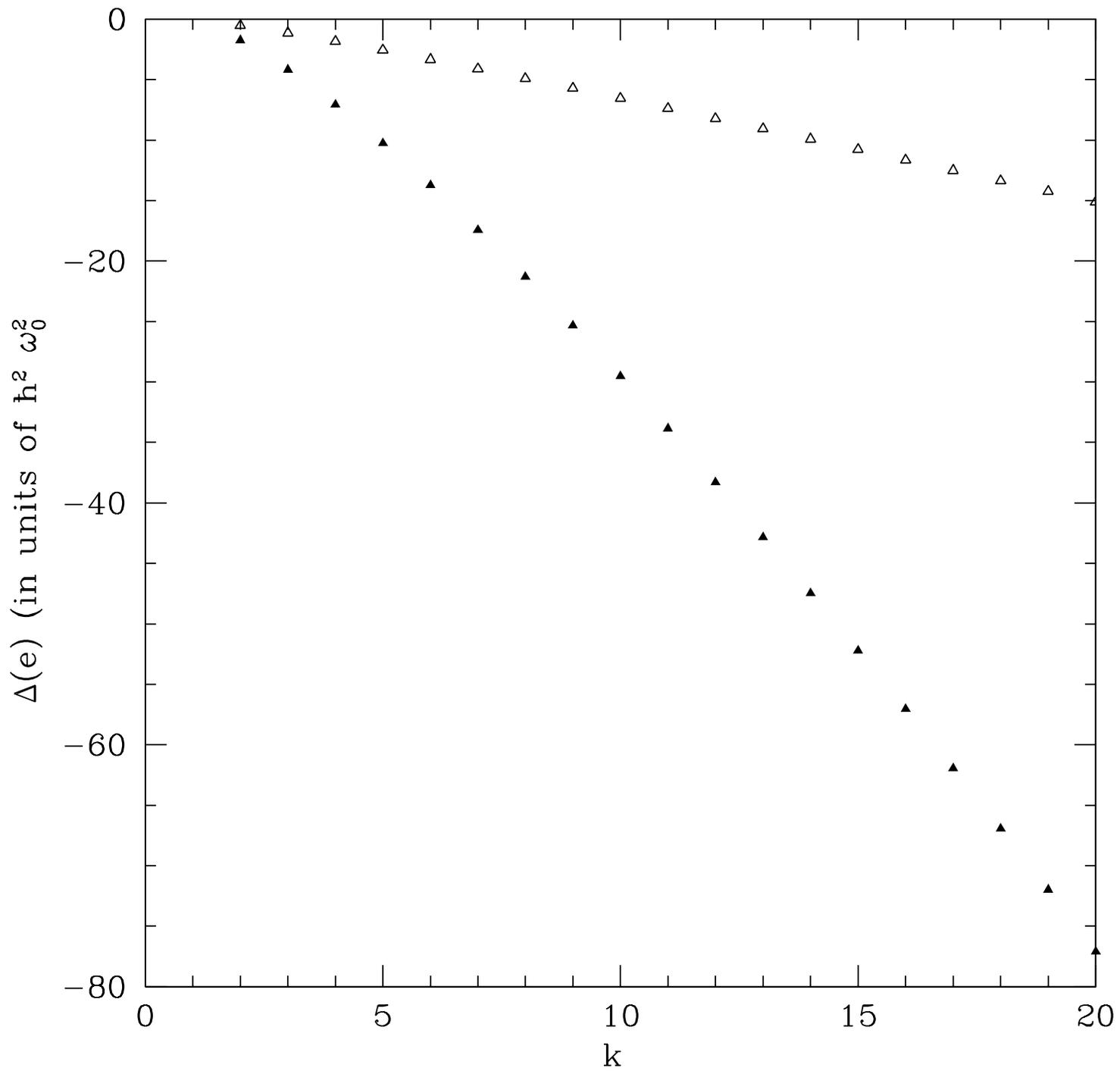
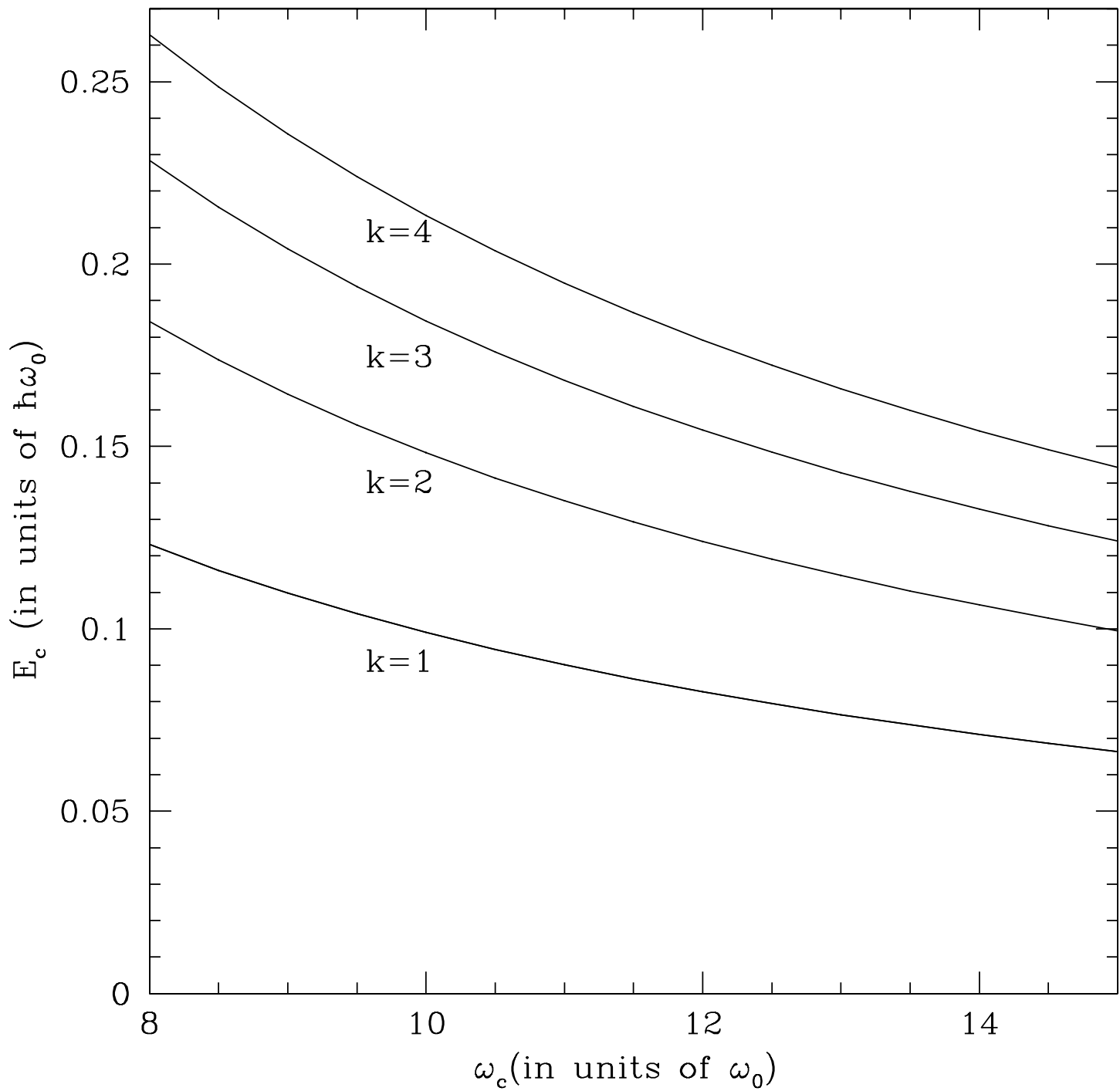
\end{document}